# Expression profiles of acute lymphoblastic and myeloblastic leukemias with ALL-1 rearrangements

T. Rozovskaia[*], O. Ravid-Amir[†], S. Tillib[‡], G. Getz[†], E. Feinstein[f], H. Agrawal[†], A. Nagler[£], E. Rappeport[±], I. Issaeva[*], Y. Matsuo[ʃ], U. R. Kees[π], T. Lapidot[+], F. Lo Coco[§], R. Foa[§], A. Mazo[‡], T. Nakamura[‡], C.M. Croce[‡#], G. Cimino[§], E. Domany[†#] and E. Canaani[*#]

Departments of [*]Molecular Cell Biology, [†]Physics of Complex Systems and [+]Immunology, Weizmann Institute of Science, Rehovot, 76100 Israel; [‡]Kimmel Cancer Center, Jefferson Medical College, Philadelphia, PA 19107; [f]Quark Biotech Inc., Cleveland OH 44106; [ʃ]Fujisaki Cell Center, Fujisaki, Okayama, Japan; [π]Institute for Child Health Research, Subiaco, Western Australia; [§]Department of Cellular Biotechnology and Hematology, University La Sapienza, Rome, 00161 Italy; [£]Sheba Medical Center, Tel-Hashomer, Israel; [±]Children Hospital of Philadelphia, PA

[#]To whom correspondence may be addressed. E-mail: croce@calvin.jci.tju.edu
fedomany@wisemail.weizmann.ac.il
eli.canaani@weizmann.ac.il

Contributed by C.M. Croce

**Summary**

The ALL-1 gene is directly involved in 5-10% of ALLs and AMLs by fusion to other genes or through internal rearrangements. DNA microarrays were utilized to determine expression profiles of ALLs and AMLs with ALL-1 rearrangements. These profiles distinguish those tumors from other ALLs and AMLs. The expression patterns of ALL-1-associated tumors, in particular ALLs, involve oncogenes, tumor suppressors, anti apoptotic genes, drug resistance genes etc., and correlate with the aggressive nature of the tumors. The genes whose expression differentiates between ALLs with and without ALL-1 rearrangement were further divided into several groups enabling separation of ALL-1- associated ALLs into two subclasses. Further, AMLs with partial duplication of ALL-1 vary in their expression pattern from AMLs in which ALL-1 had undergone fusion to other genes. The extensive analysis described here draws attention to genes which might have a direct role in pathogenesis.

Chromosome band 11q23 is a region of recurrent rearrangements in human acute leukemias. These rearrangements, usually in the form of reciprocal chromosome translocations, affect 5-10% of children and adults with acute lymphoblastic (ALL) and acute myeloblastic (AML) leukemias. The most common translocations are t(4;11) and t(9;11) accounting for 40% and 27%, respectively, of all 11q23 rearrangements. There is a strong association between leukemia phenotype and particular rearrangements. Thus, t(4;11) occurs nearly exclusively in ALL, and 85% of cases with t(9;11) are AMLs (1, 2). Essentially all 11q23 abnormalities involve the ALL-1 gene (also termed MLL, HRX or HTRX) which rearranges with more than 30 partner genes to produce fusion proteins composed of ALL-1 amino terminus and the carboxy terminus of the partner protein (3, 4). A second, and less frequent type of ALL-1 rearrangements, does not involve partner genes but rather partial duplications of ALL-1 N-terminal segments (5). ALL-1-associated leukemias show some unusual and intriguing features (reviewed in 6, 7). First, they predominate infant acute leukemias, amounting to 80% of infants with ALL and 65% of those with AML. Second, they account for the majority of therapy-related (secondary) leukemias, developing in 5-15% of primary cancer patients treated with drugs, such as etoposide (VP16) or doxorubicin (Dox), that inhibit DNA topoisomerase II. Third, in infant leukemia and in therapy-related leukemia the disease arises after a brief latency. In fact, studies of monozygotic twins and newborns with leukemia, and analysis of neonatal blood spots from children who were diagnosed with leukemia indicate that in most or all infant leukemias ALL-1 rearrangements occur *in utero*. The short latency suggests that ALL-1 fusion proteins induce leukemia with few, if any, additional mutations. Fourth, prognosis of patients with 11q23 abnormalities is dismal. Recent large studies indicated that less than 25% of infants and adults above 40 years old with ALL and t(4;11), or with AML and t(9;11) were curable (1, 2, 8).

The unique biological and clinical features of 11q23 associated leukemias, in conjunction with their induction by altered versions of ALL-1, a highly intricate chromatin modifier (9), prompted us to look for molecular clues for those features by examining the expression profiles of these leukemias.

**Materials and Methods**

**Patients and specimens.** Apart from two individuals, all patients with 11q23 abnormalities were adults. The samples were provided by GIMENA Italian Multicenter Study Group. Also included in the analysis were four AML cell lines with t(9;11) (MONOMAC6, TPH1, MOLM13 and PER377) and one with t(6;11) (ML2), and two ALL cell lines with t(4;11) – RS4;11 and B1. Genes picked up in the supervised analysis, as well as those pointed out as separating ALLs with and without t(4;11) in nonsupervised analysis, had similar expression profiles in cell lines and primary tumors. The primary tumors included 12 ALLs with t(4;11) obtained from 10 adults, one child and one infant, and 10 AMLs of adults including 5 with t(9;11), 3 with ALL-1 partial duplication and single cases of t(10;11) and t(11;19). Controls comprised of 10 AMLs of adults, 11 ALLs of adults and 2 ALLs of children. Details regarding the patients may be found in Table 2 in PNAS web site (www.pnas.org). Bone marrow samples were obtained from newly diagnosed patients.

**DNA microarray analysis.** RNAs were extracted from fresh or cryopreserved mononuclear cells by using Trizol reagent (Sigma), and assessed for their integrity by gel electrophoresis. 10 µg aliquots of total RNAs were utilized to prepare biotin-labeled cRNAs according to Affymetrix protocol. These RNAs were subsequently hybridized to human U95 oligonucleotide probe arrays corresponding to 12600 sequences (genes) (Affymetrix). Arrays were scanned and the expression value for each gene was calculated using Affymetrix software. This raw expression data was re-scaled to compensate for variations between arrays in hybridization intensity.

**Preprocessing and filtering of data.** The expression data was organized in a matrix of $n_s=52$ columns (hybridizations) and 12,600 rows (genes on the chip). Denote by $A_{gs}$ the "average difference" of gene *g* in sample $s$. First, we thresholded the data; we set $T_{gs} = A_{gs}$ for sizeable values, $A_{gs} \geq 10$, and replaced low values, $A_{gs} < 10$, by $T_{gs} = 10$. Next, log was taken: $E_{gs} = \log_2 T_{gs}$, and the genes were filtered on the basis of their variation across the samples. Denote by $\overline{E}_g$ the average of the $E_{gs}$ values obtained for gene $g$ over all $n_s$ tumor samples and by $\sigma_g$ their standard deviation. Only those genes that satisfied $\sigma_g > 1.1$ were studied. 3090 genes passed this filtering procedure; after removal of non-human Affymetrix controls and genes appearing on only one of the versions of the U95A chip, we were left with 3064 genes (out of 12,600). All further analysis was done on these genes.

**Supervised analysis.** We used supervised analysis (hypothesis testing) to identify genes, one at a time, whose expression levels can be used to separate tumours into two known classes $A$, $B$ of $n_A$ and $n_B$ samples, respectively. We used the Wilcoxon Rank Sum test in order to find genes differentially expressed between the two groups of samples (e.g. AML samples with chromosome translocations versus those without; other comparisons are listed in the Results section). We used the Rank Sum test because it is non-parametric - it does not assume normal distribution of the data. For each gene $g$ we make the null hypothesis, according to which all $n_A + n_B$ expression levels were drawn from the same distribution. The test produces a statistic $W_g$ and a p-value for each gene $g$. A p-value of $p_g = 0.05$ means that the probability of erroneously concluding that a gene does separate the two groups is 5%, which is the standard value used in the literature. However, here we deal with multiple comparisons; at this level of $p_g$ it is expected that 150 out of 3,000 random, identically distributed genes will be falsely identified as separating the two groups of samples. To control the number of false positives, we used the FDR method (10). The $N$ tested genes are ordered according to their increasing $p_g$ values and a parameter $q$, that controls the fraction of false positives is set. We then identify the minimal index $j$ such that for all $i > j$ we have $p_i \geq i \times q/N$. The null hypothesis is rejected for all genes with index $i \leq j$. This procedure yields a list of genes for which the expected fraction of false positives is $q$.

**Unsupervised Analysis: clustering.** Only unsupervised methods, such as clustering, can reveal differences that were not anticipated. We used the *Coupled Two-Way Clustering* (*CTWC*) method (11), which focuses on correlated groups of genes, one group at a time. We assume that each such group is important for one particular process of interest. Thereby the noise generated by the large majority of genes that are not relevant for that process is eliminated; furthermore, by using a group of correlated genes, noise of the individual measurements is averaged out and reduced. The relevant subsets of genes and samples are identified by means of an iterative process, which uses, at each iteration level, *stable* gene and sample clusters that were generated at the previous step. Before each clustering operation the rows of the data matrix (genes) are centered (mean=0) and normalized (standard deviation = 1). The ability to focus on stable clusters that were generated by any clustering operation is essential for the *CTWC* method; otherwise, there would be computationally unfeasible number of gene/sample cluster pairs to test (11). Since most clustering methods do not have a reliable inherent stability measure for clusters, we used **Superparamagnetic Clustering (SPC)**, a physics-based algorithm (12), that does provide a stability index, $\Delta T(C)$, to each cluster *C*. *SPC* was tested on data from a large number of problem areas including image analysis, speech recognition, computer vision and gene expression (11,12 and ref therein). A parameter *T* controls the resolution at which the data are viewed; as *T* increases, clusters break up, and the outcome is a dendrogram. A cluster *C* is "born" at $T = T_1(C)$, the value of *T* at which its "parent" cluster breaks up into two or more subclusters, one of which is *C*. As *T* increases further, to $T_2(C) > T_1(C)$, *C* itself breaks up and "dies"; $\Delta T(C) = T_2(C) - T_1(C)$ is the stability index provided by SPC. The larger $\Delta T(C)$, the more statistically significant and stable (against noise in the data and fluctuations) is the cluster *C* (13).

**Results**

**Expression profiles of ALLs with t(4;11).** Leukemic cells of ALLs with t(4;11) display features of precursor B cells with IgH rearrangements, negative for CD10 and positive for CD19, but also show some characteristics of myeloid cells (1). This and their capability to differentiate *in vitro* into monocyte-like cells had suggested that the leukemic clones originate from an early precursor cell. Hence this leukemia is classified as pro B or pre pre B. To determine whether the expression repertoire of ALLs with t(4;11) is unique we compared it to the transcription profiles of a set of ALL samples lacking t(4;11). These consisted of CD10⁻ pro B-cell ALLs, Ph chromosome-positive early pre B-cell ALLs, CD10+ early pre B-cell ALLs and T-cell ALLs. Supervised analysis indicated that at FDR (false discovery rate) of 0.05 there were 130 overexpressed and 107 underexpressed genes in ALLs with t(4;11), in comparison to ALLs lacking the abnormality (Fig. 1A). To evaluate consistency of the pattern the relative expression of each gene in all the samples was displayed in the form of bars (see examples in Fig. 1B). The top genes on the lists of overexpressed or underexpressed genes in ALLs with t(4;11) are shown in Table 1. The complete lists may be found in Table 3 as supplementary information. The clear difference in expression profiles between ALLs with the t(4;11) abnormality and other types of ALLs establish that the former belong to a unique and distinguishable class of ALL.

Examination of the genes whose expression pattern distinguishes t(4;11) ALLs from other ALLs reveals a substantial number of genes associated with growth control, cell transformation or malignancy; those genes may be classified into several functional categories:

1. overexpressed oncogenes – a) HOX A9 and MEIS1, which form a sequence specific DNA binding complex (14), are frequently co-activated in spontaneous AML of BXH-2 mice (15). Forced co-expression of the two genes in murine bone marrow cells rapidly induces AML (16). b) HOX A10, which induces AML in mice (17). c) LMO2 (RMBT2), whose overexpression, resulting from chromosome translocations, is associated with T-cell ALL (18). d) MYC, which has a critical role in cell proliferation and is deregulated in human lymphomas and other tumors (19). e) LGALS1 (galectin1), which cooperates with RAS in cell transformation (20) and whose overexpression correlates with progression of glioblastoma (21). f) PDGFRB (platelet-derived growth factor receptor beta), which is a tyrosine kinase, and is deregulated through chromosome translocations and gene fusions in chronic myeloproliferative diseases (22).

2. overexpressed genes involved in drug resistance – a) CD44, associated with aggressive B-CLL (23) and conferring resistance to several widely-used anticancer drugs (24). b) DHFR (dihydrofolate reductase), conferring resistance to methotraxate. c). BLMH (bleomycine hydrolase). d) CAT (catalase), which protects from oxidative stress.

3. overexpressed genes involved in protection from apoptosis and in survival – a) CDC2 (cell division cycle 2; p34; CDK1), which preserves the viability of cancer cells in response to microtubule poisons and anticancer drugs like vincristine and taxol, by increasing expression of the apoptosis inhibitor survivin (25). b) PPP2R5C (phosphatase 2A ), implicated in regulation of growth, transcription and signal transduction. Required for survival and protects from apoptosis in Drosophila (26). c) MAP3K5 (MAP kinase kinase kinase 5), involved in activation of the p38 MAP kinase required for initiation of the G2/M checkpoint (27), and selectively activated in non-small cell lung cancer (28).

4. underexpressed pro-apoptotic genes – a) ITPR3 (inositol 1,4,5-triphosphate receptor type 3), which mediates the release of intracellular calcium and consequently actively promotes apoptosis (29), b) IGFBP3 (IGF binding protein 3), which has proapoptotic activity both dependent and independent of p53 (30). c) JUN, implicated as positive modulator of apoptosis induced in hematopoeitic progenitor cells of the myeloid linkage (31). Downregulation of JUN might account for the failure of glucocorticoid therapy (32).

5. underexpressed tumor suppressors and growth inhibitors – a) FHIT (fragile histidine triad), target of chromosome aberrations and inactivated in many cancers, including lung, oesophagus, stomach, breast, kidney and leukemias (33). b) DAPK1 (death-associated protein kinase 1), which counters oncogene-induced transformation by activating a p19ARF/p53 apoptotic checkpoint (34). c) MADH1 (mothers aginst decapentaplegic homologue 1; SMAD1), transcription modulator mutated in various forms of cancer (35).

6. overexpressed genes acting in cell cycle progression and cell proliferation – a) CCNA1 (cyclin A1), which functions in S phase and mitosis and its expression is elevated in a variety of tumors including AMLs (36). b) BMYB (myb-like 2), which is required for proliferation of hematopoietic cells (37) and directly activates the anti-apoptotic gene ApoJ/clusterin (38). c) CDKN3 (cyclin dependent kinase inhibitor 3), which interacts with cyclin dependent kinases and is overexpressed in breast and prostate cancer (39).

Our battery of ALLs lacking t(4;11) consisted of tumors at various stages of differentiation including pre B-, pro B- and T-cell ALLs. Therefore, the differences in expression found should be due in part to the differences in differentiation stage between t(4;11) to the other ALLs. Hence we now tried to: 1) identify those genes whose expression pattern is directly correlated with the t(4;11) abnormality, either resulting from the abnormality or specifically associated with the cell type in which the chromosome translocation occurred. 2) separate the genes above from genes whose expression reflects (sensitive to) the differences between early vs late differentiation stage (pro B- vs pre B- and T-cell tumors). 3) identify genes associated with unique features of CD10$^-$ ALLs. To this end we defined three groups of ALL samples: (i) t(4;11) tumors (pro B-cells), (ii) CD10$^-$ tumors (pro B-cells), and (iii) rest of the ALLs (pre B- and T-cells). Three distinct supervised analyses were performed, which separate: 1) t(4;11) ALLs from the rest of ALLs. 2) t(4;11) ALLs from CD10$^-$ ALLs. 3) CD10$^-$ ALLs from the rest of ALLs.

The genes that participate in one or more separations were identified (see the Venn diagram of Fig. 2A). Three overlapping groups were found, containing 80, 46 and 21 genes (lists of genes in Tables 4-6 in the web site). Each of these groups contained genes that are overexpressed or underexpressed; the expression matrix of the three groups is shown in Fig. 2B. 80 genes separate both pro B-cell t(4;11) ALLs from pre B- and T-cell ALLs, as well as the latter from pro B-cell CD10$^-$ ALLs. Having been picked in both separations, this group of 80 genes distinguishes pro B-cell ALLs [both with and without the t(4;11) chromosome translocation] from pre B- and T-cell ALLs. The 46 genes of the second intersection separate simultaneously t(4;11) ALLs from CD10⁻ ALLs and from pre B- and T-cell ALLs. Being singled out in both separations, this group of genes is neither associated with the differences between pro B- vs pre B- and T-cells and tumors, nor does it involve specific features of CD10⁻ tumors. Rather, the expression of these 46 genes is affected directly by the t(4;11) abnormality and probably by other unique features of the pro B-cells in which the t(4;11) aberration occurred. The majority of these 46 genes also appear in Table 1. The last group of 21 genes separates CD10⁻ from t(4;11) ALLs as well as from pre B- and T-cell ALLs. Being selected in both separations, these 21 genes are likely to be associated with unique features of CD10⁻ tumors.

Inspection of Fig. 2B points to three t(4;11) tumors, samples 2, 6, 14, which show a variant transcription profile. While the expression pattern of the 46 genes, specifically correlated with t(4;11) ALLs, is similar in these three tumors and in the rest of t(4;11) ALLs (see genes 81-126 of Fig. 2B), the transcription profile of the three tumors with regard to genes 1-80 (which distinguish pro B- from pre B- and T-cell tumors) is closer to pre B- and T-cell ALLs, unlike the profile of the other eleven t(4;11) samples. The three tumors also show some quantitative variation from the other t(4;11) ALLs in transcription of the genes whose expression is associated with CD10⁻ ALLs (genes 127-147) (Fig. 2B). These results suggest the existence of two sub-families of ALLs with the t(4;11) chromosome translocation, distinguished by their expression patterns.

Finally, we applied the coupled two-way clustering method (11, 12) in an unsupervised analysis. A group of 25 genes was found consistently underexpressed in ALLs with t(4;11) compared to the other ALLs (Fig. 3; Table 7 in the web site). The cluster of samples with low expression of these genes includes 13/14 of t(4;11) and 3/4 of CD10⁻ ALLs. This is consistent with the close similarity in biological and clinical features between these two types of tumors. A second group of 132 genes separated the seven cell lines included in the analysis from the forty-five primary tumors. All these genes were underexpressed in the cell lines (Fig. 6 and Table 8 in the web site).

**Transcription profile of AMLs with ALL-1 rearrangements.** AMLs with 11q23 translocations and ALL-1 rearrangements were compared in their expression profiles to AMLs with normal karyotypes. At FDR of 0.15 (85% confidence) we identified 67 genes overexpressed or underexpressed in AMLs with 11q23 abnormalities (Fig. 4; Table 9 in web site). Three primary AMLs with ALL-1 partial duplication (5) were compared to the other AMLs with regard to expression of the 67 genes. Two of the three tumors resembled AMLs without 11q23 abnormalities, while the third appeared closer to the tumors with chromosome translocations (Fig. 4). The similarity between AMLs without 11q23 aberrations and AMLs with ALL-1 partial duplications was further evidenced in the failure to separate the two groups at an acceptable FDR. (In parallel, AMLs with 11q23 abnormalities were separated from AMLs with ALL-1 partial duplications at FDR of 0.3; some of this analysis is shown in Fig. 7 and Table 10 in the web site). These results, if confirmed with additional samples, suggest molecular variations between AMLs triggered by recombination of the ALL-1 gene to partner genes and AMLs triggered by ALL-1 partial duplications. The variations might be reflected in biological and clinical features.

Examination of the list of genes most correlated with AMLs carrying 11q23 abnormalities (Table 9 in the web site) discloses some involved in cancer or related processes. These include the overexpressed insulin receptor which enhances DNA synthesis and inhibits apoptosis (40), the overexpressed repair gene RAD 51 which is upregulated in breast and pancreatic cancers (41) and probably increases drug resistance, the overexpressed PPP2R5C phosphatase, the underexpressed JUNB which upregulates the tumor suppressor gene p16 and represses cyclin D1 (42) and whose knockout in mice induces myeloproliferative disease (43), the underexpressed tumor suppressor FHIT, the underexpressed double stranded RNA-activated protein kinase proapoptotic PRKR, which upregulates FAS and BAX (44), and the underexpressed DEFA1 (defensin) involved in immune response.

Having identified genes differentially expressed in ALLs with t(4;11) compared to ALLs without t(4:11), and in AMLs with 11q23 abnormalities compared to AMLs without such abnormalties, we intersected the results of these two tests (we used FDR level of 0.15 for both) in order to find the genes in common. We identified 52 such genes that were overexpressed or downregulated in the relevant tumors (Fig. 5 and Table 11 in the web site). For all these genes the difference was high for one type of tumors (e.g. ALLs), but modest for the second type (e.g. AMLs). The genes that were overexpressed in the samples with ALL-1 rearrangements included the phosphatase PPP2R5C, and the MCM4 gene whose product is an essential component of the prereplicative complex (45). The underexpressed genes included FHIT and JUNB.

**Discussion**

Our results indicate distinct transcription profiles of ALL-1 associated tumors. This is likely to be reflected in the unusual clinical and biological characteristics of these tumors, such as short latency, poor prognosis, expression of myeloid genes in ALL, etc. Some of the genes pinpointed in our study of ALLs with t(4;11), which were mostly adults, were also indicated (Table 3) in our previous preliminary analysis (46) and in recent investigations which dealt with ALLs from infants and children (47, 48).

Examining the genes overexpressed or underexpressed in tumors with ALL-1 rearrangements (in particular in ALLs) indicates constellation of expression patterns previously associated with, and/or highly favorable for malignant transformation and cancer. This includes activation of oncogenes (MYC, HOX A9 and MEIS1, LMO2, etc.), inactivation of tumor suppressor genes such as FHIT and DAPK1, suppression of apoptosis by downregulation of pro-apoptotic genes and upregulation of survival genes, suppression of host immune response (upregulation and downregulation of galectin 1 and defensin, respectively), upregulation of genes such as CD44, DHFR and bleomycin hydrolase conferring drug resistance, and overexpression of genes involved in cell prolferation (e.g. cyclin A1 and myb-like 2). Some of the overexpressed genes we identified, like VLDL, PDGFRB, HOX A9, MEIS1 and insulin receptor, are also found expressed in normal hematopoetic stem cells (49) but the majority of genes are not. We suggest that at least some of the genes alluded to by our study contribute directly to the aggressive nature of the disease and to its known resistance to therapy.

In an attempt to identify genes whose expression correlates more strictly with the t(4;11) genotype, we separated away genes which distinguish pro B- from pre B- and T-cell tumors. The resulting list of 46 genes (Table 5) includes several oncogenes and tumor suppressor genes and probably constitutes a better database from which to choose genes for further experiments. Another approach taken to identify genes more likely to be associated with the pathogenesis was based on the assumption (still unproven) that ALL-1 fusion proteins trigger malignancy by a similar mechanism in both ALLs and AMLs. Thus, we looked for genes which behave in similar fashion (upregulated or downregulated) in ALLs and AMLs with ALL-1 rearrangements (Fig. 5). At the top of the list we find PPP2R5C, FHIT and JUNB.

Compartmentalization into two groups of the genes whose expression distinguishes t(4;11) from other ALLs resulted in the unexpected identification of two subclasses of t(4;11) tumors (Fig. 2B). The subclasses are discerned by the expression profile of the 80 genes separating pro B- from pre B- and T-cell tumors. Since t(4;11) tumors are generally considered pro B-cell ALL, it is surprising that with regard to genes separating pro B- from pre B- and T-cell ALLs, the smaller subclass of t(4;11) appears close to pre B- and T-cell tumors. Comparison of the clinical records of the corresponding two subclasses of patients (Table 2; samples ht17, 21 and 27 in this table show the variant profile) indicates that in the first group there are 2/3 long-term survivors, but in the second group the outcome is worse (2/9). How wide-spread is the distribution of t(4;11) patients into two groups and whether there is a significant correlation with survival remains to be determined. Finally, clinical studies have shown (8) that infants younger than 1 year, with ALL and t(4;11), fared significantly worse than older children. It will be of interest to compare these two groups of tumors with regard to the expression of the two groups of genes (46 and 80) identified here.

The supervised analysis of AMLs with ALL-1 rearrangements vs control AMLs showed a less uniform pattern, as well as a lower number of separating genes. This suggests that the two groups of tumors are more heterogeneous. Surprisingly, two of the three AMLs with ALL-1 partial duplications showed expression profiles resembling AML controls. The generality of this observation should be decided by analyzing additional tumors.

**Acknowledgements**

O. R-A. thanks Amnon Amir for helpful thoughts and ideas. These studies were supported by NCI grant CA 50507 and by grants from the Israel Academy of Science, US-Israel BSF, Israel Cancer Research Fund, Minerva Foundation and the Germany-Israel Science Foundation (GIF).

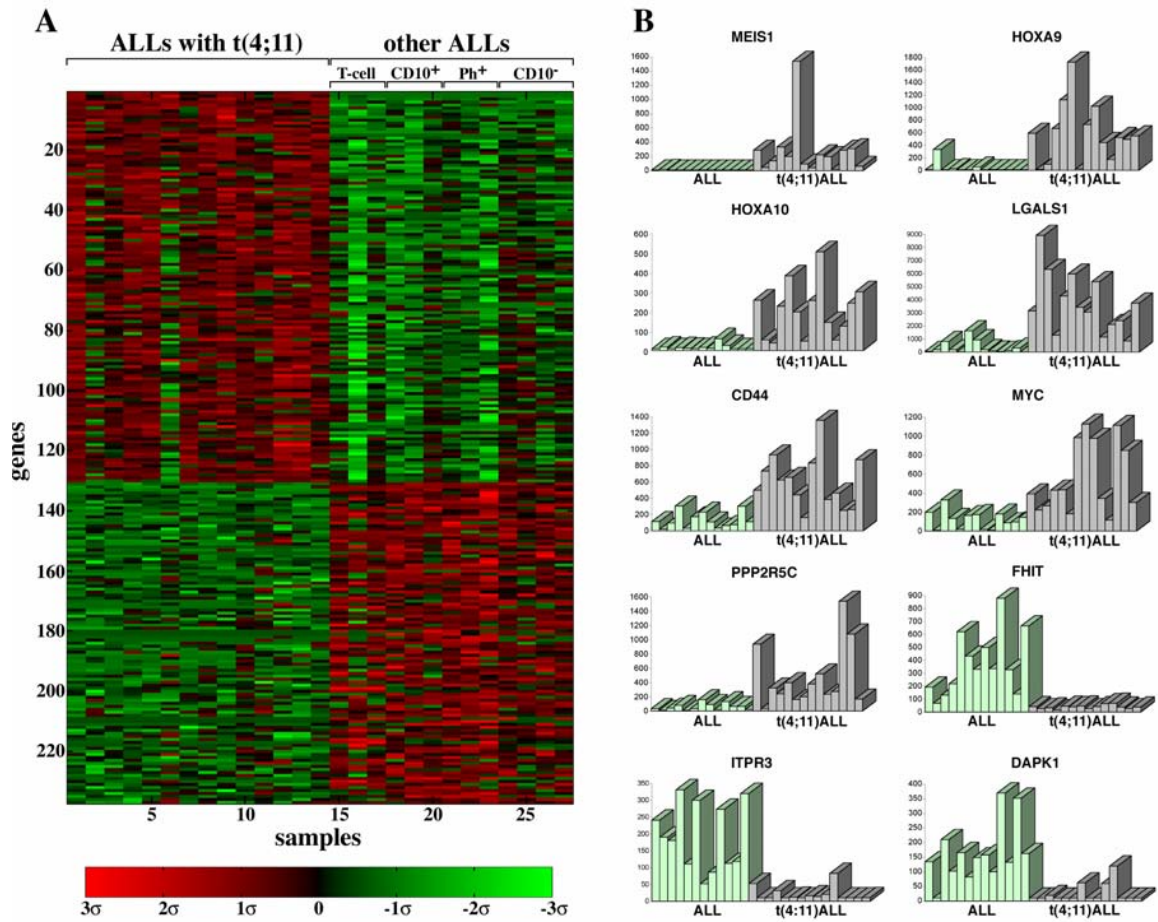

**Fig. 1** Supervised analysis of genes distinguishing ALLs with ALL-1 rearrangement [t(4;11)] from other ALLs (A), and relative levels of expression of selected genes (B). Expression levels greater and smaller than the mean 0 are shown in red and green, respectively.

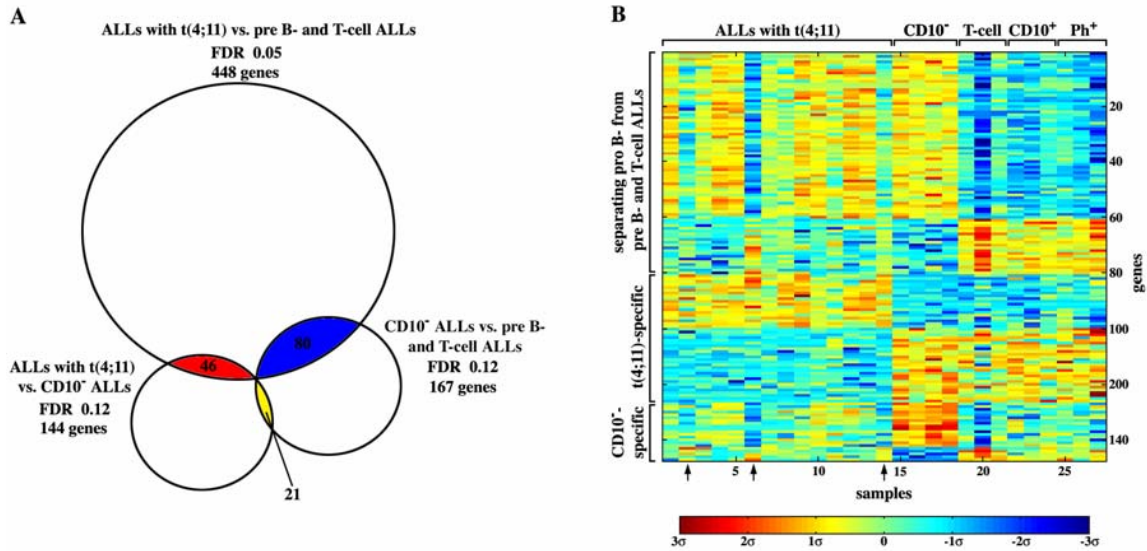

**Fig. 2** Intersections of genes separating three types of ALLs (see text). Three groups of genes, encompassing 80, 46 and 21 genes, were found to participate each in two separations (A). The expression matrix of these three groups is shown in (B). Levels of expression higher or lower than the mean 0 are shown in red/yellow and blue, respectively. Arrows point to samples with variant expression profile (see text).

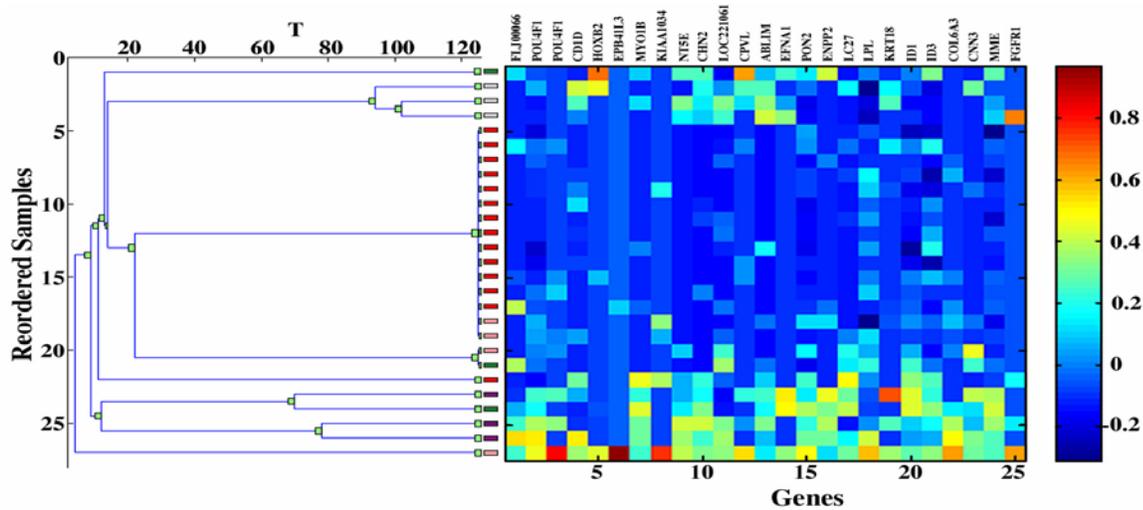

**Fig. 3** Clustering the ALL samples, on the basis of their expression levels over a cluster of 25 genes, G7 (that was obtained by the CTWC method). The resulting dendrogram is on the left; each leave corresponds to an ALL sample, with t(4;11) ALLs colored red and CD10⁻ ALLs by rose. The expression matrix is shown on the right, with rows corresponding to samples and columns to genes. 13/14 of t(4;11) samples and 3/4 of CD10⁻ ALLs are in the central cluster of samples with low expression levels.

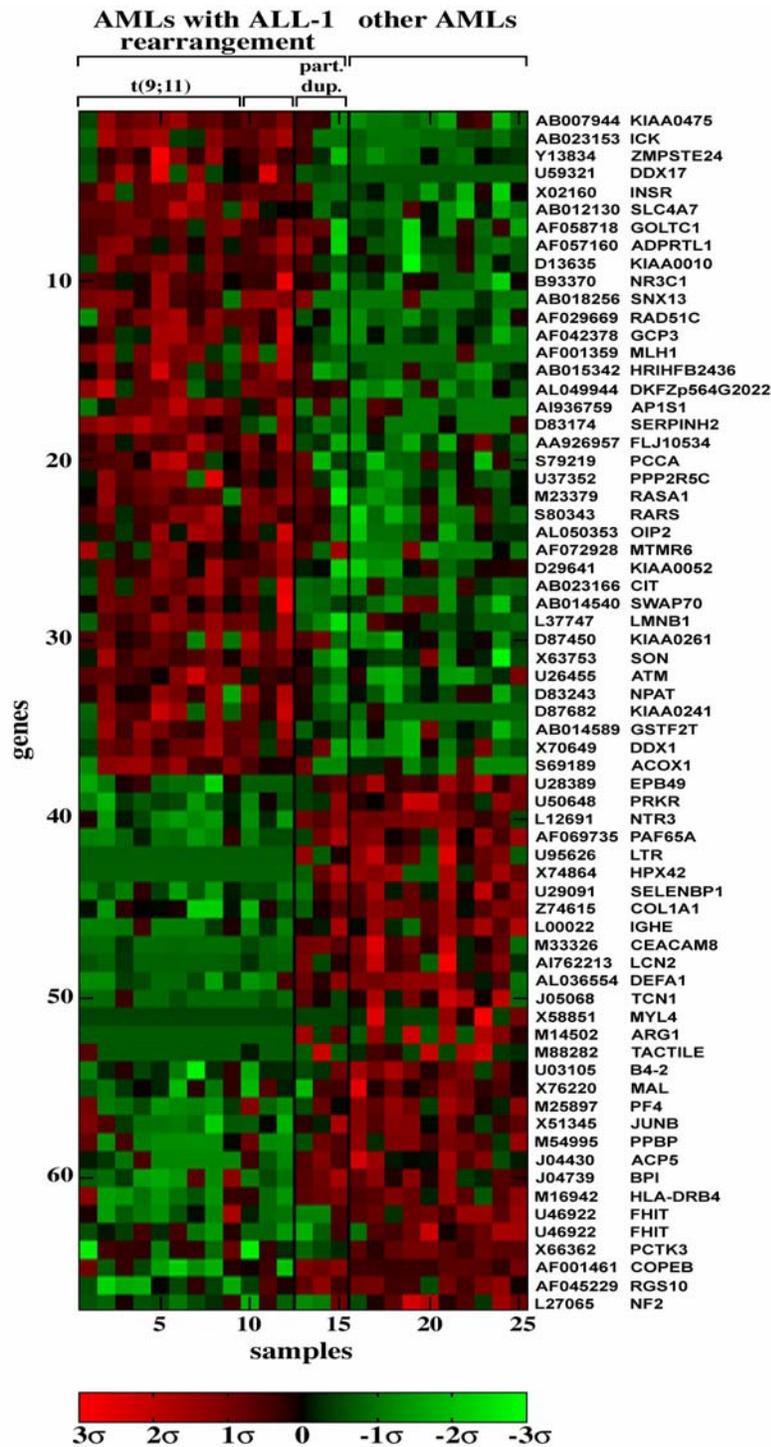

**Fig. 4** Genes distinguishing AMLs with 11q23 chromosome translocations and ALL-1 rearrangements (samples 1-12) from other AMLs (samples 16-25). Samples 13-15 of AMLs with ALL-1 partial duplication were not included in the supervised analysis, but were added later for purpose of comparison.

**TABLE 1. GENES MOST CORRELATED (OVEREXPRESSED OR UNDEREXPRESSED) IN ALLs WITH THE t(4;11) ABERRATION, COMPARED TO OTHER ALLs**

**OVEREXPRESSED**

| No | Also scored as t(4;11)-specific* | Acc. No | Symbol and function | p-value | fold change | confidence interval |
|---|---|---|---|---|---|---|
| 1 | √ | D16532 | VLDLR, very low density lipoprotein receptor; signal transduction, modulation of Dab1/Tau phosphorylation | 0.000004 | 17.51 | (10.67-28.74) |
| 2 | √ | U85707 | MEIS1, myeloid ecotropic viral integration site 1 homolog (mouse); homeobox protein, murine leukemia | 0.000004 | 14.50 | (7.64-27.51) |
| 3 | √ | AC004080 | HOXA10, homeo box A10; sequence specific transcription factor | 0.000022 | 10.80 | (6.17-18.90) |
| 4 | √ | AI535946 | LGALS1, lectin, galactoside-binding, soluble 1 (galectin1); cell apoptosis and differentiation | 0.000024 | 23.00 | (8.82-59.98) |
| 5 | √ | M54992 | CD72 antigen; B cell proliferation and differentiation | 0.000037 | 4.35 | (2.80-6.74) |
| 6 | √ | U41813 | HOXA9, homeo box A9; sequence specific transcription factor, murine myeloid leukemia | 0.000041 | 20.12 | (7.22-56.09) |
| 7 |  | AF098641 | CD44, CD44 isoform (Indian blood group system) | 0.000056 | 4.43 | (2.65-7.41) |
| 8 | √ | L05424 | CD44, CD44 antigen (Indian blood group system); cell surface receptor, lymphocyte activation/homing | 0.000056 | 5.43 | (3.07-9.60) |
| 9 | √ | AA099265 | RECK, reversion-inducing-cystein-rich protein with kazal motifs; membrane glycoprotein, tumor suppression | 0.000063 | 3.58 | (2.03-6.31) |
| 10 | √ | M14087 | HL14, beta-galactoside-binding lectin | 0.000068 | 6.94 | (3.50-13.76) |
| 11 | √ | Z69030 | PPP2R5C, protein phosphatase 2, regulatory subunit B (B56), gamma isoform | 0.000069 | 7.55 | (3.60-15.83) |
| 12 | √ | M59040 | CD44, CD44 antigen (Indian blood group system); cell surface receptor, lymphocyte activation/homing | 0.000069 | 4.10 | (2.57-6.54) |
| 13 |  | D83767 | D8S2298E (reproduction 8); fertilization | 0.000086 | 3.16 | (2.01-4.97) |
| 14 |  | AF016004 | GPM6B, glycoprotein M6B; integral membrane protein, expressed in neurons and glia | 0.000095 | 13.06 | (5.98-28.53) |
| 15 |  | X96753 | CSPG4, chondroitin sulfate proteoglycan 4 (melanoma-associated) | 0.000097 | 7.97 | (3.56-17.85) |
| 16 | √ | D78177 | QPRT, quinolinate phosphoribosyltransferase; biosynthesis of NAD and NADP | 0.000104 | 7.31 | (3.86-13.86) |
| 17 | √ | V00568 | MYC, v-myc myelocytomatosis viral oncogene homolog (avian); transcription factor, cell proliferation | 0.000126 | 5.93 | (2.68-13.12) |
| 18 |  | X61118 | LMO2, LIM domain only 2 (rhombotin-like 1); transcription factor, red blood cell development | 0.000126 | 3.85 | (2.05-7.22) |
| 19 |  | Z69030 | PPP2R5C, protein phosphatase 2, regulatory subunit B (B56), gamma isoform | 0.000153 | 6.11 | (2.79-13.38) |
| 20 | √ | M58597 | FUT4, fucosyltransferase 4 (alpha (1,3) fucosyltransferase, myeloid-specific); glycosylation | 0.000187 | 3.52 | (2.17-5.71) |

**UNDEREXPRESSED**

| No | Also scored as t(4;11)-specific* | Acc. No | Symbol and function | p-value | fold change | confidence interval |
|---|---|---|---|---|---|---|
| 131 | √ | U46922 | FHIT, fragile histidine triad; nucleotide metabolism, tumor suppressor | 0.000010 | -8.18 | (-5.16)-(-12.97) |
| 132 | √ | U70321 | TNFRSF14, tumor necrosis factor receptor superfamily, member 14; lymphocyte activation | 0.000012 | -24.73 | (-12.05)-(-50.72) |
| 133 | √ | U01062 | ITPR3, inositol 1,4,5-triphosphate receptor type 3; signal transduction, small molecule transport | 0.000013 | -10.69 | (-6.49)-(-17.63) |
| 134 | √ | M16594 | GSTA2, glutathione S-transferase A2 | 0.000017 | -3.48 | (-2.27)-(-5.34) |
| 135 | √ | U03858 | FLT3LG, fms-related tyrosine kinase 3 ligand; stimulates proliferation of early hematopoietic cells | 0.000024 | -2.24 | (-1.56)-(-3.20) |
| 136 | √ | AB007895 | KIAA0435 | 0.000037 | -4.38 | (-2.47)-(-7.77) |
| 137 |  | J05257 | DPEP1, dipeptidase 1 (renal); renal metabolism of glutathione | 0.000046 | -3.24 | (-2.07)-(-5.07) |
| 138 | √ | X53586 | ITGA6, integrin alpha 6; laminin receptor, critical structure role in the hemidesmosome | 0.000056 | -15.57 | (-6.59)-(-36.79) |
| 139 | √ | J03600 | ALOX5, arachidonate 5-lipoxygenase; biosynthesis of leukotrienes | 0.000056 | -4.57 | (-2.63)-(-7.94) |
| 140 |  | U01062 | ITPR3, inositol 1,4,5-triphosphate receptor, type 3; signal transduction, small molecule transport | 0.000062 | -5.21 | (-3.30)-(-8.23) |
| 141 | √ | L34059 | CDH4, cadherin 4, type 1, R-cadherin (retinal); cell adhesion | 0.000069 | -6.89 | (-3.48)-(-13.64) |
| 142 | | AF041434 | PTP4A3, protein tyrosine phosphatase type IVA, member 3 | 0.000085 | -4.11 | (-2.36)-(-7.18) |
| 143 | √ | X76104 | DAPK1, death-associated protein kinase 1; mediating interferon-gamma-induced cell death | 0.000093 | -7.90 | (-3.94)-(-15.85) |
| 144 | √ | U19969 | TCF8, transcription factor 8; repression of transcription (interleukin-2) | 0.000100 | -3.29 | (-1.85)-(-5.84) |
| 145 | √ | M92357 | TNFAIP2, tumor necrosis factor, alpha-induced protein 2 | 0.000104 | -3.82 | (-2.20)-(-6.66) |
| 146 | √ | U46922 | FHIT, fragile histidine triad; nucleotide metabolism, tumor suppressor | 0.000104 | -3.76 | (-2.04)-(-6.91) |
| 147 | √ | AB002301 | KIAA0303 | 0.000110 | -15.21 | (-6.31)-(-36.68) |
| 148 | √ | S59184 | RYK, RYK receptor-like tyrosine kinase | 0.000135 | -4.64 | (-2.58)-(-8.33) |
| 149 | | Y11312 | PIK3C2B, phosphoinositide-3-kinase, class 2, beta polypeptide | 0.000187 | -2.52 | (-1.66)-(-3.81) |
| 150 | | M22324 | ANPEP, alanyl (membrane) aminopeptidase (aminopeptidase N, M, CD13, p150); receptor for coronavirus | 0.000216 | -4.24 | (-2.23)-(-8.06) |
| 151 | √ | U59423 | MADH1, mothers against decapentaplegic homolog 1 (Drosophila); receptor-regulated transcription | 0.000253 | -10.52 | (-4.20)-(-26.35) |

**\* Also included within the group of 46 genes, associated with specific features of t(4;11) ALLs, in Fig. 2**